# Investigating the Impacts of Generative AI on Information Seeking

Alexi Orchard
Technology and Digital Studies
University of Notre Dame
Notre Dame, IN, USA
aorchard@nd.edu

Shannon Lodoen
Humanities and Communication
Embry Riddle Aeronautical University
Prescott, AZ, USA
lodoens@erau.edu

## ABSTRACT

This paper is an encore submission of our 2026 journal article "Expertise and Information Seeking in the Age of Generative AI: New Procedures, New Problematics" with an extended discussion for the CSCW 2026 "Broader Impacts of GenAI in Communication" Workshop on October 10, 2026. In the original article, we employ procedural rhetoric to analyze how generative AI chatbots leverage natural language signifiers of expertise and intelligence to influence users' perception of their trustworthiness. In this submission, we extend our conversation in the CSCW community with the goal of cultivating a cross-disciplinary vocabulary for describing, analyzing, and mitigating the risks posed by the integration of generative AI into human communication practices. It is important to develop an understanding of how the procedures surrounding information-seeking practices are informed by users' values, experiences, and expectations—and how these procedures might in future be altered by the emerging turn towards AI "experts" and authority.

## 1 Introduction

The process of gathering, evaluating, and applying new information is a shared human experience that goes back millennia. As information has been codified and collected in new ways (such as libraries, archives, websites, and databases), methods of information seeking have, accordingly, developed in tandem. Routine information seeking has historically involved the consultation of reputable sources, whether that be the local newspaper, a family doctor, or a trusted neighbor. While traditional forms of information seeking are now well understood, the advent of generative AI has introduced new complexities and considerations for information seeking in many different contexts. Indeed, for both scholars and the broader public, information seeking is being drastically reconfigured, in particular by the emergence of generative AI chatbots.

Generative AI chatbots, built on language model (LM) architectures, have begun to fulfill a variety of roles [1] for human users, ranging from schedulers and virtual assistants to tutors, coaches, counsellors, and confidantes. Indeed, Hirvonen et al. assert that "artificial intelligence is poised to reshape human behavior, with its influence extending to search engines, social media, and nearly every aspect of the Internet" as summarized in [2]. If search engines are already conceived of as "cognitive authorities" and "influential sources of expertise" [3], what does this shift to generative AI portend for questions of expertise, authority, and information seeking? These questions are being taken up across disciplines beyond AI and computing, including law, politics, medicine, and the humanities [4], [5], [6], [7], [8].

Looking specifically at the intersection of rhetoric (our field of study) and generative AI, research has been largely limited to analyzing, producing, or testing various prompts or generated outputs. Topics of investigation include "Socratic" chatbots for pedagogical purposes [9], [10], [11], concerns over perceptions of accuracy and truthfulness [12], [13], [14], and testing of various LMs' capabilities to produce rhetorically effective, i.e. persuasive, content [15], [16], [17], [18], [19]. Aiming to expand this content-specific focus, in previous work [20], [21], we described the procedural (i.e., non-content based) means through which chatbots persuade users to reconfigure their thoughts, behaviors, and practices in three domains: information seeking, language construction, and social relationship building. In this paper, we continue in this vein of thought with an explicit focus on information-seeking procedures. Specifically, we examine the process of seeking, accessing, and utilizing "expert" information [21]. We are, again, not focusing on the content of AI-generated responses so much as on how information is presented for users to interact with and the patterns of engagement that are established. The ability for laypeople to accurately recognize and act on expertise is crucial across many contexts, but perhaps most especially in the act of information seeking. The seemingly authorless, near-instantaneously generated text has, it seems, its own aura of expertise that convincingly persuades users of the veracity of the information it conveys.

By combining scholarly literature and research across computer science, science communication, and rhetorical studies, we highlight how humanities-based scholarship can inform ethical analyses of emerging technologies. As such, we address a multi-disciplinary audience that includes not only researchers, scholars, and educators, but also the everyday users of generative AI chatbots who are most implicated in, and impacted by, the changes wrought by AI in the information-seeking landscape.

## 2 Key Concepts

### 2.1 Procedural Rhetoric

While not commonly applied in STEM-focused work, the concept of rhetoric is highly useful, especially as it pertains to the notion of expertise [22]. Broadly construed, the term refers to "the art of persuasion," effective communication, or symbolic expression. Stretching from the fourth century BCE to present day, studies in and of rhetoric seek to identify what Aristotle termed the "means of persuasion" available in myriad situations and environments [23]. These means of persuasion can be found not only in "human-to-human" rhetoric (i.e., in speech or writing), but also through the modulation and manipulation of public space, visuals, objects, and processes.

*Procedural* rhetoric, as coined by Ian Bogost, refers to "the art of persuasion through rule-based representations and interactions rather than the spoken word, writing, images, or moving pictures" [24]. A procedural rhetorical lens considers how users construct and find meaning in their interactions with processes or procedures, which may range from computer programming to government bureaucracies [24, p. 5]. Crucially, procedural rhetoric holds that processes can create arguments in favor of certain practices or modes of interaction. These nonverbal, indirect arguments are formulated through the choices that are made available to a participant; the steps that are set out as "necessary" or "optional"; and the level of challenge or ease in achieving the end goal. These modulable elements can "persuade" a participant of the validity of certain ideas, beliefs, or actions. Procedural rhetoric is, accordingly, found in the procedures that govern or structure information-seeking practices, such as the layouts of libraries or specific protocols for handling archive materials. Such "procedures" shape how users find and interpret information, and how they can mobilize or utilize that information. Human-computer interactions (i.e., how a user interacts with their AI chatbot) are also examples of structured, procedural arguments—which emerge, for instance, in the types of options users have for engaging or interacting with the interface. When AI chatbots respond to human users' queries, they are following a series of encoded procedures that determine why and how their responses will be formed [20]. When users turn to AI chatbots for responses to their queries, this also creates a new set of procedures to study rhetorically, which is what we aim to do with this research.

As Bogost emphasizes, procedures are important because they "found the logics that structure behavior" [24, p. 7]; to study procedures is to study the values, beliefs, goals, and assumptions that underpin them. In asking "how does this work?" when analyzing "cultural, social, and historical systems," we are really concerned with understanding the "logics that motivate their human actors" [24, p. 8]. While seemingly authorless, AI-generated text nonetheless reflects human motives, values, and ideals that have been embedded within it, as they are in all technological objects, systems, and structures. Accordingly, there is a need to understand how the procedures surrounding information-seeking practices are informed by users' values, experiences, and expectations—and how this might in future be altered by the emerging turn towards AI "experts" and authority. As more people use chatbots, they will not only reference but may come to rely on LMs for information previously obtained through a human source.

### 2.2 Information Seeking

Since the mid-twentieth century, information-seeking practices have been examined across multiple disciplines, beginning in library and information science [25] and later extending to psychology, health and science communication, and information studies [26], [27], [28]. According to Kuhlthau, information seeking refers to "user's constructive activity of finding meaning from information in order to extend his or her state of knowledge on a particular problem or topic" [29, p. 361]. Generally speaking, information seeking involves finding and evaluating resources, thereby determining if they are relevant and satisfactory for the task at hand. The formal process of conducting research—whether by browsing the stacks, querying a database, or searching the web (not to mention earlier forms of oral traditions or epistolary networks, for example)—has always been characterized by the available technology and the cultural assumptions of what it means to know something. These modes of information seeking have evolved with and through emerging technology. Generative AI is the latest innovation in a long history of technology used for this purpose—though, as Lund et al. note, "the roots of artificial intelligence in information seeking can be traced back to the 1950s, when researchers began exploring the use of computers for processing and evaluating data" [2]. Some of the new opportunities that have emerged in information seeking through the integration of AI systems include: "multimodal search, information retrieval in human-like interaction, and creation and modification of information to meet specific information needs" [30, p. 1159]. Generative AI represents a notable shift from Google search, though it is still misperceived as doing information retrieval. A brief comparison of these information-retrieval methods illuminates the distinct processes through which knowledge is constructed, accessed, and obtained. Typing "headache remedies" into the Google search bar will yield many pages of possible sources, which can be altered through changes to the search query. Conversely, typing "how can I get rid of my headache?" into an AI chatbot will elicit a human-like exchange, starting with something to the effect of: "Here are some effective ways to get rid of headaches, depending on the type and severity. . ." The user can then further clarify their query: “I think I have a tension headache” or “What causes cluster headaches?” or “Can I take Advil with a migraine?” As the user provides more specific prompts, the chatbot delivers increasingly precise responses. While both methods have their practical advantages and disadvantages, we are interested in the rhetorical implications of how these methods shape users’ experience with accessing, interpreting, and integrating information.

## 2.3 Expertise

The term “expert,” as Ericsson explains, is “derived from the same root as experience and experiment, which refers to efforts to learn from experience”; “expertise” refers, generally, to the state of having “gained special skills or knowledge representing mastery of a particular subject through experience and instruction” [31, p. 508]. Accordingly, as Collins and Evans describe, gaining expertise and “becoming an expert” are processes that require both individual accomplishment and external recognition [32].

Despite the wealth of academic literature on expertise, however, the concept continues to come under fire in popular discourse. Indeed, the question of what constitutes expertise—and who or what can claim expert status—has come to the forefront in recent political conversations, the COVID-19 pandemic, and now, increasing interest in LMs and AI chatbots. If expert status is “achieved by a complex negotiation with an audience of epistemic claims, a cultivation of skills, and capacities for deliberation and moral judgment” [22, p. 3], how do we account for the status of generative AI systems that do not engage in any of these practices? While AI systems do engage in a problem-solving process similar to that of human information seekers (as described in [2] and [33]), this does not mean that the AI system can itself be considered an expert, or that its outputs should be accepted unquestioningly.

# 3 Relevance to the CSCW Community

When the key concepts above coalesce, they lead us to some challenging questions: What effect will the new AI-enabled procedures for information seeking have on users (and knowledge construction writ large)? What does it mean to trust a chatbot with information seeking, to imbue it with the authority to find, filter, and assemble information? What happens when decision-making is outsourced to a chatbot that cannot have true experience or direct knowledge in the world it claims to understand? These questions are a central stake in the future of information seeking and communication practices more broadly. The CSCW workshop [34] highlights relevant examples to consider here.

Whether working in tandem with human expert consultation or independently, generative AI users are shifting away from the search-and-recall method wherein the onus is on the user to find, consume, and comprehend information of varying levels of difficulty, relevance, and reliability. Instead, the user asks questions and is rewarded with information that has already been located, scanned for relevance, summarized, and delivered in a highly readable and accessible format. The valuable process of gathering and sifting through sources, which itself supports the ability to discern relevance and credibility, is reduced or eliminated altogether. Meanwhile, the user’s understanding of the given topic is systematically narrowed as they receive popular or generic information, while more diverse and innovative perspectives may go unnoticed unless the user specifically knows how to prompt for such anomalies. The way that LMs respond to user queries, inputs, and ideas is instrumental in constructing the user’s experience of interacting with and learning from the system. As discussed in prior work, this shift in information seeking behavior may have implications for a user’s ability to develop critical thinking skills, identify misinformation and bias, and assess the veracity of information in high stakes scenarios [35], [36], [37], [38].

To close with a specific use case for generative AI in information seeking: in one case, an AI assistant designed to generate simplified summaries of judicial opinions was shown to help non-experts understand key features of a ruling [5]. In situations where specialized jargon is the norm, such as legal, scientific, or medical contexts, it may be reasonable of non-experts to seek accessible, non-judgmental, or otherwise comfortable situations not perceived to be associated with a human expert. In AI chatbot interactions, users receive (and perhaps prefer) “an almost human explanation” [39]. This “almost human” version is devoid of the discomforts potentially associated with the typical rules of engagement with a legal or medical professional, which may involve opaque jargon or unclear resolutions. In this way, AI chatbots circumvent previously necessary pathways to information and can thereby reshape users’ expectations and experiences with technical or specialized knowledge.

As generative AI chatbots become even more common in everyday life, the key terms we have highlighted here—rhetoric, expertise, and information seeking—are liable to shift, develop, and arise in new situations. We hope that they may provide a relevant and enlightening springboard for studying the impacts and emerging risks of generative AI-mediated information-seeking behaviors. As Bogost writes: “processes influence us. They seed changes in our attitudes, which in turn, and over time, change our culture” [24, p. 340]. Understanding the processes users engage in is therefore crucial for taking stock of the ways in which generative AI influences users’ perceptions, perspectives, and practices.